\documentstyle[12pt,epsfig]{article}

\begin{document}
\begin{titlepage}
\vspace{-2mm}
\rightline{gr-qc/yymmnnn}

\vskip 1cm

\centerline{\large {\bf Electric Charge in Interaction with
Magnetically Charged}}

\centerline{\large{\bf Black Holes}}

\vspace{1.5mm}

\vspace{10mm} \centerline{J.H. Kim$^\dagger$\footnote{E-mail:
rtiger$@$phya.snu.ac.kr} and Sei-Hoon
Moon$^\star$\footnote{E-mail: jeollo@add.re.kr}}

\vskip 5mm

\centerline{$^\dagger${\it Center for Theoretical Physics $\&$
School of Physics and Astronomy }}

\centerline {{\it College of Natural Sciences, Seoul National
University, Seoul 151-742, Korea }}

\centerline{$^\star${\it WETV Group, Center of Joint Modeling and
Simulation, Agency for Defense Development }}

\centerline {{\it P.O.Box 35 Yuseong, Daejeon 305-600, Korea}}

\vskip 15mm


\begin{abstract}
We examine the angular momentum of an electric charge $e$ placed
at rest outside a dilaton black hole with magnetic charge $Q$. The
electromagnetic angular momentum which is stored in the
electromagnetic field outside the black hole shows several common
features regardless of the dilaton coupling strength, though the
dilaton black holes are drastically different in their spacetime
structure depending on it. First, the electromagnetic angular
momentum depends on the separation distance between the two
objects and changes monotonically from $eQ$ to $0$ as the charge
goes down from infinity to the horizon, if rotational effects of
the black hole are discarded. Next, as the black hole approaches
extremality, however, the electromagnetic angular momentum tends
to be independent of the distance between the two objects. It is
then precisely $eQ$ as in the electric charge and monopole system
in flat spacetime. We discuss why these effects are exhibited and
argue that the above features are to hold in widely generic
settings including black hole solutions in theories with more
complicated field contents, by addressing the no hair theorem for
black holes and the phenomenon of field expulsion exhibited by
extremal black holes.
\end{abstract}

\vspace{5mm}

\indent\indent\hspace{4mm}  Keywords: Angular momentum, dilaton
black hole, monopole.


\end{titlepage}

\newpage

\section{Introduction}\label{I}

The influence of a gravitational field on external fields around
black holes was investigated in 1970s and 1980s by a number of
authors. In particular, the electrostatic field near a black hole
was derived by Hanni and Ruffini \cite{HR} for a point charge at
rest, by Cohen and Wald \cite{CW} for multipole fields, and by
Linet \cite{Linet} in algebraic form in Schwarzschild spacetime.
Leaute and Linet \cite{LL} applied this test field approximation
to the case of a Reissner-Nordstr\"{o}m black hole background.
Recently, the problem of a point charge at rest near a
Reissner-Nordstr\"{o}m black hole was reconsidered by Bini,
Geralico and Ruffini \cite{BGR,BGR1} using the first order
perturbation approach formulated by Zerilli \cite{Ze}. External
stationary magnetic fields were also considered in the presence of
Kerr black holes using the test field approximation by Wald
\cite{RMW} and of Reissner-Nordstr\"{o}m black holes employing the
first order coupled perturbations by Bi$\check{\rm c}\acute{\rm
a}$k and Dvo$\check{\rm r}\acute{\rm a}$k \cite{BD}. The authors
found that, as electromagnetic multipole moments approach the
horizon of a black hole, all the multipole moments fade away
except the monopole. That is, the no hair conjecture for the
electromagnetic multipole moments holds for the black hole and so
there is no black hole with multipole moments other than the
monopole. Another noticeable result of these works is that the
component of the electromagnetic field normal to the horizon tends
to vanish as the black holes approach extremality and the flux
lines are totally expelled in the extremal limit. This is
analogous to the ``Meissner effect" for a magnetic field around a
superconductor \cite{CEG,AF}.

In this paper we consider a system of an electric charge and a
magnetically charged dilaton black hole (instead of an
electrically charged Reissner-Nordstr\"{o}m black hole). We don't
try to derive the electric field structure of the charge. Instead,
we examine the angular momentum associated with the system. As is
well known, the angular momentum of a static configuration of an
electric charge $e$ and a magnetic monopole $Q$ is precisely $eQ$
in flat space \cite{Js}. The angular momentum is all stored in the
electromagnetic fields. However, when the monopole is replaced
with a magnetically charged black hole, the electromagnetic
angular momentum\footnote{We will use this term to designate the
angular momentum stored in the electromagnetic field outside the
black hole.} changes in a somewhat complicated way. Such system
was considered originally by Garfinkle and Rey \cite{SJR} many
years ago and by Bunster and Henneaux \cite{CBMH} recently. As was
shown by Garfinkle and Rey, for a system of an electric charge $e$
held at rest outside a magnetically charged black hole of strength
$Q$ in the Einstein-Maxwell theory, the angular momentum stored in
the electromagnetic field outside the black hole does depend upon
the separation distance between the two objects. The
electromagnetic angular momentum changes monotonically from $0$ to
$eQ$ as the charge moves from the horizon to infinity. However, in
the extreme limit of the black hole it does not depend on the
separation and is precisely $eQ$, as in the system of an electric
charge and a monopole in flat space.

We shall see below that these results remain to be true even when
the theory includes dilatonic couplings. This is somewhat striking
because charged black hole solutions of the dilaton gravity
\cite{GM,GHS} have many properties distinct from those of the
Reissner-Nordstr\"{o}m black hole, depending on the strength of
the dilaton coupling $\alpha$. Each dilaton black hole has only
one horizon and a spacelike singularity giving rise to a
Schwarzschild-type spacetime structure, whereas the typical
Reissner-Nordstr\"{o}m black hole has two horizons and a timelike
singularity. Furthermore, the event horizon of a dilaton black
hole is actually singular in the extremal limit and has zero area,
as opposed to that of the Reissner-Nordstr\"{o}m black hole which
has finite area and is nonsingular. The thermodynamical
relationships also differ depending on the dilaton coupling
$\alpha$ which is a nonnegative constant. In the extremal limit
the dilaton black holes have zero entropy and their temperature is
zero for $\alpha<1$, finite for $\alpha=1$, and infinite for
$\alpha>1$. Extremal black holes for $\alpha>1$ act more like
elementary particles than usual Reissner-Nordstr\"{o}m black holes
because infinite potential barriers form outside the horizon
\cite{HW}. Many other features of the dilaton black hole solutions
(but not all) depend on the coupling strength $\alpha$. In spite
of these distinctive differences, however, the qualitative
features of the electromagnetic angular momentum are not modified
in the presence of the magnetically charged dilaton black hole
(instead of the Reissner-Nordstr\"{o}m black hole). Regardless of
the strength of the dialton coupling, the electromagnetic angular
momentum also changes monotonically from $0$ to $eQ$ as the charge
moves from the horizon to infinity for non-extreme dilaton black
holes, while it is independent of the separation distance with the
precise value $eQ$ in the extremal limit of the black hole.

Since the dilaton black holes are drastically different in their
spacetime structure from the Reissner-Nordstr\"{o}m black holes,
the fact that qualitative features of the electromagnetic angular
momentum are independent of the dialton coupling makes us expect
that such features are common to black holes in various theories
with more complicated field contents. Indeed, the nature of the
electromagnetic angular momentum seems to be generic for various
black holes as far as the no hair theorem holds for the black
holes and electromagnetic fields are expelled from the horizon in
the extremal limit of the black holes. We shall argue that the
separation dependence of the electromagnetic angular momentum in
the non-extremal case is related to the nontrivial spacetime
structure, {\it i.e.}, the existence of the non-degenerate
horizon, and the no hair nature of black holes. On the other hand,
the fact that it is independent of the separation distance and its
value is precisely $eQ$ in the extremal limit comes from the
phenomenon of flux expulsion exhibited by extreme black holes
mentioned in the first paragraph of this section.

The plan of our work is as follows. In the next section, we shall
first summarize the magnetically charged black hole solution to
the Einstein-Maxwell-dilaton gravity. In section \ref{III}, we
derive an equation governing the angular momentum function of the
system of an electric charge and a magnetic dilaton black hole by
extending Garfinkle and Rey's work. In section \ref{IV}, we
analyze solutions of the equation. In section \ref{V}, we examine
the electromagnetic angular momentum of the charge and black hole
system analytically and numerically. In the last section, we argue
that the monotonic variation of the electromagnetic angular
momentum from 0 to $eQ$ in the non-extremal black hole case comes
from the no hair nature of the black hole and that its
independence from the separation distance in the extremal limit of
black holes results from the flux expulsion by extremal black
holes.

\section{Magnetically Charged Dilaton Black Holes}\label{II}

In this section we give a brief introduction to the charged
dilaton black holes in $3+1$ dimensions found by Gibbons {\it
et~al.}\cite{GM} and Garfinkle {\it et~al.}\cite{GHS} and slowly
rotating dilaton black holes discussed by Horne {\it
et~al.}\cite{HH}. The action for the Einstein-Maxwell-dilaton
gravity is given by, in the Einstein frame,
\begin{eqnarray}\label{action}
S=\int d^4x\sqrt{-g}\left[-{\rm R}+2(\nabla \phi)^2+e^{-2\alpha
\phi}F^2\right],
\end{eqnarray}
where ${\rm R}$ is the scalar curvature, $\phi$ the dilaton field
and $F_{\mu\nu}$ the Maxwell field strength. $\alpha$ is a
non-negative constant governing the strength of the dilaton
coupling to the Maxwell field. For $\alpha=0$ this action
corresponds to the standard Einstein-Maxwell action with an extra
free scalar field. When $\alpha=1$, this action describes a part
of the tree-level low energy limit of string theory in the
Einstein frame. The case $\alpha=\sqrt{3}$ gives simply the
Kaluza-Klein action which is obtained by dimensionally reducing
the five-dimensional vacuum Einstein action. The equations of
motion derived from the above action (\ref{action}) are
\begin{eqnarray}
&&\nabla_a (e^{-2\alpha \phi}F^{ab}) = 0, \label{eom1} \\
&&\nabla^2 \phi+ \frac{\alpha}{2}~ e^{-2\alpha \phi}F^2 = 0,
\label{eom2}  \\
&&{\rm R}_{ab}=2\nabla_a\phi\nabla_b\phi+2 e^{-2\alpha \phi}
F_{ac}{F_b}^c-\frac12~g_{ab} e^{-2\alpha\phi} F^2.\label{eom3}
\end{eqnarray}

The magnetically charged and spherically symmetric static black
hole solution of the above equations takes the form \cite{GM,GHS}
\begin{eqnarray}
F&=&Q\sin\theta~d\theta \wedge d\varphi,\label{emf}\\
e^{-2\phi}&=& \left(1-\frac{r_-}{r}\right)^{2\alpha/(1+\alpha^2)},
 \label{df}\\
ds^2&=& -\lambda^2 dt^2 + \frac{dr^2}{\lambda^2} + R^2 d\Omega^2,
 \label{dbh}
\end{eqnarray}
where
\begin{eqnarray}
\lambda^2 &=& \left(1-\frac{r_+}{r}\right) \left(1-\frac{r_-}{r}
\right)^{(1-\alpha^2)/(1+\alpha^2)},\label{lamsq}\\
R^2&=& r^2 \left(1-\frac{r_-}{r}\right)^{2\alpha^2/(1+\alpha^2)}
\label{R2}.
\end{eqnarray}
Here, $r_+$ and $r_-$ are the values of the parameter $r$ at the
outer and the inner horizon, respectively, and are related to the
physical mass and the magnetic charge by
\begin{eqnarray}
M &=& \frac{r_+}{2}+\left(\frac{1-\alpha^2}{1+\alpha^2}\right)
\frac{r_-}{2},\label{mpm} \\
Q &=&  \left(\frac{r_+ r_-}{1+\alpha^2}\right)^{1/2}\label{qpm}.
\end{eqnarray}
$R$, not $r$, has the normal meaning of the radial variable, in
the sense that the area of the sphere obtained by varying $\theta$
and $\varphi$ at fixed $t$ and $r$ (or $R$) is $4\pi R^2$.

The qualitative features of dilaton black holes depend crucially
on $\alpha$. When $\alpha=0$, this solution reduces to the
Reissner-Nordstr\"{o}m solution of the Einstein-Maxwell theory.
The geometry is then not singular at either of $r_-$ and $r_+$.
For all $\alpha$, the surface $r=r_+$ is an event horizon.
However, for $\alpha\neq 0$ the geometry is singular at $r_-$,
because $R$ vanishes there. In fact, the surface $r=r_-$ is a
spacelike curvature singularity very similar to that in the
Schwarzschild metric. The singularity is shielded by the horizon
if $M\geq M_c\equiv Q/\sqrt{1+\alpha^2}$. Due to this difference,
there occurs a big difference in the spacetime structure of the
extremal black holes in the two cases. For $\alpha=0$ the area of
the event horizon is finite. However, for $\alpha\neq 0$ the area
vanishes for the extremal black holes and the geometry is singular
there.

The temperature and entropy of the black holes are given by,
respectively,
\begin{eqnarray}
T&=&\frac{1}{4\pi r_+}\left(1-\frac{r_-}{r_+}\right)^{(1-\alpha^2)/
(1+\alpha^2)},\\
S&=&\pi {r_+}^2\left(1-\frac{r_-}{r_+}\right)^{2\alpha^2/
(1+\alpha^2)}.
\end{eqnarray}
Evidently the extremal black holes will have finite entropy in
case $\alpha=0$, but zero entropy otherwise. The temperature of
the extremal black holes is zero for $\alpha<1$, finite and equal
to $1/8\pi M$ for $\alpha=1$, and infinite for $\alpha>1$.
Extremal black holes with non-zero temperature inevitably develop
naked singularities and are therefore not physically acceptable.
However, Holzhey and Wilczek \cite{HW} showed that black holes
with $\alpha>1$ have infinite mass gaps. These black holes can
neither absorb any finite energy from impinging objects nor
radiate into these modes. In these senses the $\alpha>1$ extremal
black holes act more like elementary particles.

Now we brief on the rotating dilaton black hole solution for
further works in the following sections. Since there is no
explicit solution describing rotating charged black hole for
arbitrary $\alpha$, the rotating solution is considered in the
limit of slow rotation of \cite{HH}. $g_{t\phi}$ is the only term
in the metric that changes to first order in the angular momentum
parameter $a$. The dilaton does not change to $O(a)$, and $A_t$ is
the only component of the vector potential that changes to $O(a)$.
For arbitrary $\alpha$, in the limit of slow rotation the metric
has the following form:
\begin{eqnarray}\label{srbh}
ds^2 = -\lambda^2(r) dt^2 + \frac{dr^2}{\lambda^2(r)} + R^2(r)
d\Omega^2-2a f(r)R^2(r)\sin^2\theta dt d\phi,
\end{eqnarray}
where
\begin{eqnarray}
f(r)& \equiv &\frac{(1+\alpha^2)^2}{(1-\alpha^2)(1-3\alpha^2)}
\frac{1}{r_-^2}\nonumber\\
& & -\left[\frac{1}{r^2}+
\frac{(1+\alpha^2)^2}{(1-\alpha^2)(1-3\alpha^2)}\frac{1}{r_-^2}
+\frac{1+\alpha^2}{1-\alpha^2}\frac{1}{r_-r}-\frac{r_+}{r^3}\right]
\left(1-\frac{r_-}{r}\right)^{(1-3\alpha^2)/(1+\alpha^2)}.
\end{eqnarray}
The surface gravity and the area of the event horizon do not
change to $O(a)$. However, for $\alpha>1$, a small amount of
rotation produces a large change in the geometry close to the
horizon of a nearly extremal black hole. In the extremal limit, as
$r$ approaches $r_+=r_-$, $f(r)R^2(r)$ diverges for $\alpha>1$.
The angular momentum changes to $O(a)$ as discussed in \cite{HH}.
The angular momentum $J$ is related to the asymptotic form of the
metric by
\begin{eqnarray}
g_{t\varphi}\sim -\frac{2J}{r}\sin^2\theta + O\left(\frac{1}{r^2}
\right),
\end{eqnarray}
which gives
\begin{eqnarray}\label{srbham}
J=\frac{a}{6}\left(3r_++\frac{3-\alpha^2}{1+\alpha^2}r_-\right).
\end{eqnarray}

\section{Equation for the Angular Momentum}\label{III}

In this section we derive an equation governing the angular
momentum of a particle of mass $m$ and charge $e$ at rest in the
field of a magnetically charged dilaton black hole. We assume that
the charge and mass of the particle is small enough and hence we
can calculate the angular momentum to first order in $m$ and $e$.
It is straightforward to solve Einstein-Maxwell-dilaton equations
Eqs. (\ref{eom1})-(\ref{eom3}) to first order, because most of
components of the metric and the gauge field change only to second
order. Though many of the interesting physical quantities come
into second order, we can still extract some useful information on
the angular momentum from the first order analysis.

To zeroth order the fields and the metric are described by Eq.
(\ref{emf})- (\ref{dbh}). Since the system is axisymmetric, the
components of the metric and the gauge field that come into first
order can be deduced from the inspection of axially symmetric
black holes like the Kerr-Newman solution. The only terms of the
metric and the gauge field that change to first order are
$g_{t\varphi}$ and $A_t$. The existence of $A_t$ also follows from
the Bianchi identity $\nabla_{[a}F_{bc]}=0$ and the fact that the
Lie derivative of $F_{ab}$ with respect to $(\partial/\partial
t)^a$ vanishes. The corrections to other components of the fields
and the metric come into second order, and can be ignored in the
present analysis. Note that the equations of motion for the
dilaton receive no correction to first order and thus the behavior
of the dilaton is unchanged from Eq. (\ref{df}). Therefore, we
need only to develop a first order equation in $g_{t\varphi}$ and
$A_t$ to get information for the angular momentum. We do this by
extending straightforwardly Garfinkle and Rey's work for a
magnetically charged Reissner-Nordst\"{o}m black hole to the case
of a magnetically charged dilaton black hole.

Let the point electric charge be at rest at a point $r=b$ on the
polar axis $\theta=\pi$ in the field of a magnetically charged
dilaton black hole, then its stress-energy tensor and current
vector are given by
\begin{eqnarray}
T^{ab}&=&\frac{m}{2\pi R^2(b) \sin\theta}~\lambda(b)^{-1}
~\delta\left(r-b\right)
\delta\left(\theta-\pi\right)\left(\frac{\partial}{\partial
t}\right)^a
\left(\frac{\partial}{\partial t}\right)^b,\\
j^a&=&\frac{e}{2\pi R^2(b) \sin\theta}~\delta\left(r-b\right)
\delta\left(\theta- \pi\right)\left(\frac{\partial}{\partial
t}\right)^a.\label{crt}
\end{eqnarray}
The system is stationary and axisymmetric, and thus has its
corresponding Killing vectors $(\partial/\partial t)^a$ and
$(\partial/\partial\phi)^a$. In an axisymmetric, asymptotically
flat spacetime with the axial Killing vector
$(\partial/\partial\phi)^a$, the total angular momentum\cite{wald}
$J$ can be calculated by performing the Komar integral defined by
\begin{eqnarray}\label{Lft}
J\equiv \frac{1}{16\pi} \int_{S} \varepsilon_{abcd}\nabla^c
\left(\frac{\partial} {\partial\varphi}\right)^d,
\end{eqnarray}
where $S$ is a two-sphere in the asymptotic region where $T_{ab}$
is assumed to vanish. This total angular momentum of the system
contains both the intrinsic spin of the black hole and the angular
momentum stored in the electromagnetic field. To extract directly
the information on the angular momentum from the equations of
motion, we define the angular momentum contained within a
two-sphere $S(r)$ of constant $r$ and $t$:
\begin{eqnarray}\label{Lft}
{\cal L}(r)\equiv\frac{1}{16\pi}\int_{S(r)}\varepsilon_{abcd}
\nabla^c \left(\frac{\partial}{\partial\varphi}\right)^d,
\end{eqnarray}
and the total angular momentum $J$ is then given by
\begin{eqnarray}\label{ttL}
J= \lim_{{r\rightarrow\infty}}{\cal{L}}(r).
\end{eqnarray}

Obtaining the equation governing ${\cal L}(r)$ is the purpose of
this section. ${\cal L}(r)$ can be expressed in terms of
$g_{t\varphi}$ and unperturbed metrics of the black hole. By
straightforwardly calculating the integral of Eq.(\ref{Lft}), we
obtain the following expression for the angular momentum to first
order in $g_{t\varphi}$
\begin{eqnarray}\label{Lft1}
{\cal L}(r)
=\frac{1}{6}R^4\frac{d}{dr}\left(\frac{\chi}{R^2}\right),
\end{eqnarray}
where $R^2$ is the radial metric component of (\ref{R2}) and
$\chi$ is defined as follows
\begin{eqnarray}\label{chi}
\chi\equiv\frac34\int^\pi_0 g_{t\varphi}\sin\theta~d\theta.
\end{eqnarray}
Since $g_{t\varphi}$ and $A_t$ appear mixed in the Einstein and
Maxwell equations, it will be convenient to define another
quantity $u$ related to $A_t$ by
\begin{eqnarray}\label{ur}
u\equiv-\frac32\int_0^\pi A_t \sin \theta \cos \theta d\theta .
\end{eqnarray}

Taking the $t-\varphi$ component of Einstein equation (\ref{eom3})
and the $t$ component of Maxwell equation (\ref{eom1}), evaluating
all quantities to first order in $g_{t\varphi}$ and $A_t$, and
using the definitions (\ref{chi}) and (\ref{ur}), we find the
following equations for $u$ and $\chi$:
\begin{eqnarray}
\lambda^2\frac{d^2\chi}{dr^2}-\frac{2}{R^2}\left(1-\frac12
\frac{dR^2}{dr}\frac{d\lambda^2}{dr}+Q^2\frac{e^{-2\alpha\phi}}
{R^2}\right)\chi+4Q\frac{e^{-2\alpha\phi}}{R^2}u=0,\label{Ein}\\
\frac{\lambda^2}{R^2}\frac{d}{dr}\left(R^2\frac{du}{dr}\right)
+\lambda^2\frac{r^2}{R^2}\frac{d}{dr}\left(\frac{R^2}{r^2}\right)
\frac{du}{dr}-\frac{2u}{R^2}+\frac{2Q}{R^4}\chi=6\pi
e^{2\alpha\phi} \int_0^\pi j_t \sin\theta\cos\theta d\theta
\label{Max} .
\end{eqnarray}
Inserting Eqs.(\ref{df}), (\ref{lamsq}) and (\ref{Lft1}) into
Eq.(\ref{Ein}), $u$ can be expressed in terms of ${\cal L}$ as
follows
\begin{eqnarray}
u=-\frac{3}{2Q}\frac{r^2}{R^2}\lambda^2\frac{d{\cal L}}{dr}
+\frac{Q}{R^2}\chi.
\end{eqnarray}
Finally, inserting this expression into Eq.(\ref{Max}), we get
the following equation which governs the angular momentum function
${\cal L}(r)$:
\begin{eqnarray}\label{goveq}
\frac{d}{dr}\left[\frac{R^4}{r^2}\frac{d}{dr}
\left(\frac{\lambda^2 r^2}{R^2}\frac{d{\cal L}}{dr}\right)
-2\left(1+\frac{2Q^2}{r^2}\right){\cal L}\right] &=&-\frac{4\pi
QR^2}{\lambda^2}\int_0^\pi j_t \sin\theta\cos\theta
d\theta\nonumber\\
&&\nonumber\\
&=&-2eQ~\delta(r-b),
\end{eqnarray}
where we have used Eq.(\ref{crt}) to evaluate the integral.
Eq.(\ref{goveq}) is the main result of this section. It is easy to
check that, when $\alpha=0$, Eq.(\ref{goveq}) reduces to the
equation for ${\cal L}$ obtained in Ref.\cite{SJR}.

\section{Solutions of the Second-order Equation}\label{IV}

Since the right hand side of Eq.(\ref{goveq}) vanishes when $r\neq
b$, it will be convenient to consider the nonhomogeneous, linear,
second-order differential equation
\begin{eqnarray}\label{sndeq}
{\cal D}{\cal L}=C,
\end{eqnarray}
where $C$ is an integration constant and ${\cal D}$ is the
differential operator defined as
\begin{eqnarray}
{\cal D}\equiv\frac{R^4}{r^2}\frac{d}{dr}\left(\frac{\lambda^2
r^2}{R^2} \frac{d}{dr}\right)-2\left(1+\frac{2Q^2}{r^2}\right).
\end{eqnarray}
It follows from Eq.(\ref{goveq}) that $C$ is discontinuous at
$r=b$ and the difference is given by
\begin{eqnarray}\label{DZb}
C_>-C_<=-2eQ
\end{eqnarray}
where $C_>$ and $C_<$ are values of ${\cal D}{\cal L}$ in regions
$r>b$ and $r<b$ respectively. The constants $C_>$ and $C_<$ may
depend on $b$ but not on $r$.

The general solution of the second-order equation (\ref{sndeq}) is
composed of two homogeneous solutions and a particular solution,
and satisfies automatically the third-order equation (\ref{goveq})
when $r\neq b$. A particular solution $\ell_p(r)$ of the
inhomogeneous equation is given by
\begin{eqnarray}\label{lpr}
\ell_p(r)=\frac{a}{6}\left(3r_++\frac{3-\alpha^2}{1+\alpha^2}r_--
\frac{4}{1+\alpha^2}\frac{r_+r_-}{r}\right),
\end{eqnarray}
with the inhomogeneous term
\begin{eqnarray}\label{ccc}
C=-\frac{a}{3}\left(3r_++\frac{3-\alpha^2}{1+\alpha^2}r_-\right).
\end{eqnarray}
This inhomogeneous solution behaves well in the whole region of
our interest, from $r_+$ to $\infty$. As we will see in the next
section, this corresponds to the angular momentum of the slowly
rotating dilaton black hole.

On the other hand, we have no explicit solutions of the
second-order homogeneous equation (\ref{sndeq}) with $C=0$ for
arbitrary $\alpha$. However, the properties of two linearly
independent solutions of the homogeneous equations can be inferred
by examining the indicial equations near the singularities at
$r=r_+$ and $\infty$. For non-extremal black holes (as $r_+>r_-$),
the indicial equations say that one solution approaches a finite
value, but the other solution diverges in the order of
$\ln(r-r_+)$ at $r=r_+$. As $r$ goes to the infinity, {\it i.e.},
$r\rightarrow\infty$, one solution diverges in the order of $r^2$
but the other solution converges to zero in the order of $r^{-1}$.

Furthermore, it can be easily shown that there can be no
homogeneous solution finite at both the singularities, i.e.,  one
solution which is finite at $r_+$ should diverge at $\infty$ and,
on the contrary, the other solution which is divergent at $r_+$
should be zero at $\infty$. The proof is as follows: let's suppose
that there is a solution $\ell(r)$ finite at both singularities.
We know that then the solution must satisfy $\ell(r_+)\neq 0$ and
$\ell(\infty)=0$ from the indicial equations at $r_+$ and $\infty$
and we can assume that $\ell(r_+)>0$ without loss of generality.
Now it will be more convenient to rewrite the second-order
homogeneous equation (\ref{sndeq}) with $C=0$ as follows
\begin{eqnarray}
(r-r_+)(r-r_-)\ell^{\prime\prime}+\left(r_++\frac{1-3\alpha^2}
{1+\alpha^2}r_--2\frac{1-\alpha^2}{1+\alpha^2}\frac{r_+r_-}{r}
\right)\ell^\prime-2\left(1+\frac{2}{1+\alpha^2}\frac{r_+r_-}
{r^2}\right)\ell=0.
\end{eqnarray}
We easily see, from this equation, that $\ell^\prime$ is positive
at $r_+$ because the first term of the left hand side vanishes,
the coefficient of $\ell^\prime$ is positive there, and the
coefficient of $\ell$ is negative definite. On the other hand,
according to the maximum and minimum value theorem, $\ell$ must
have a maximum value at a point $r_m$ between $r_+$ and $\infty$.
At the point $r=r_m$, $\ell^\prime$ should be zero by our
assumption. However, this leads us to conclude that
$\ell^{\prime\prime}$ should be positive, and so $\ell(r)$ is
downwardly convex at $r_m$. However, this contradicts our
assumption that the function $\ell(r)$ has the maximum at $r_m$.
Thus, we can conclude there can be no solution of the homogeneous
equation (\ref{sndeq}) which is finite simultaneously at $r=r_+$
and $\infty$.

Let's suppose that $\ell_1(r)$ is such a solution of the
homogeneous differential equation (\ref{sndeq}) which is finite at
$r_+$ and divergent according to $r^2$ as $r\rightarrow\infty$.
Then, the second solution $\ell_2(r)$ can be generated as
\begin{eqnarray}\label{l12}
\ell_2(r)=\ell_1(r)\int_{r}^{\infty}\frac{R(r^\prime)^2}
{{r^\prime}^2\lambda(r^\prime)^2\ell_1(r^\prime)^2}~dr^{\prime}.
\end{eqnarray}
Evidently, $\ell_2(r)$ diverges by order $\ln (r-r_+)$ near
$r=r_+$ because $\ell_1$ is finite there, and $\ell_2$ converges
by order $r^{-1}$ as $r$ approaches $\infty$ because $\ell_1$
diverges by order $r^2$ there. It will be sufficient to know these
properties of the homogeneous solutions to extract qualitative
aspects of the angular momentum due to electromagnetic fields.

On the other hand, for extremal black holes the indicial equation
near $r=r_+$ differs from that above, and says that a solution
goes to zero according to $(r-r_+)^{D_+}$ and the other solution
diverges as $(r-r_+)^{D_-}$, where
$D_{\pm}\equiv(3\alpha^2-1\pm\sqrt{17\alpha^4+26\alpha^2+25})
/2(1+\alpha^2)$. The indicial equation at $\infty$ is not
different from that in the case of non-extremal black holes. Then,
$\ell_1(r)$ is zero at $r_+$ and diverges at $\infty$, but
$\ell_2(r)$ is divergent at $r_+$ and vanishes as
$r\rightarrow\infty$.

\section{Electromagnetic Angular Momentum}\label{V}

In this section we examine the angular momentum of an electric
charge at rest in the field of a magnetically charged black hole
using the properties of the solution shown in the previous
section. In general, the total angular momentum $J$ contains
contributions from both the rotation of the black hole itself and
the electromagnetic field around the black hole. There are
additional electromagnetic fields induced by the rotation of the
black hole besides the electric field of the charge and the
magnetic field of the black hole. Those induced fields also
contribute to the angular momentum of the system. However, we will
separate out this contribution together with the intrinsic spin of
the black hole from the total angular momentum leaving only the
portion irrespective of the rotation of the black hole, {\it
i.e.}, the `electromagnetic angular momentum' coming from the
electric field of the charge and the magnetic field of the black
hole. It will be interesting to compare this electromagnetic
angular momentum with the angular momentum of the charge-monopole
system in flat space. Probably the easiest way to find the
electromagnetic angular momentum is to compute the angular
momentum for an electric charge at rest at a distance from a
non-rotating magnetically charged black hole. However, it may not
be possible to set up such a configuration from the outset, when
the charge is held at a finite distance from the hole. The black
hole may not stay static under the influence of the electric field
of the charge, but it may keep on rotating as long as the charge
is held at a fixed position forcefully.

To see this, let's suppose a magnetically charged black hole
immersed in the weak, asymptotically uniform electric field. This
situation is equivalent, by electromagnetic duality, to the case
of an electrically charged black hole immersed in the weak,
asymptotically uniform magnetic field considered by Bi$\check{\rm
c}\acute{\rm a}$k and Dvo$\check{\rm r}\acute{\rm a}$k in
\cite{BD}. Therefore, we can directly write down the metric
perturbation from their work. For the general stationary and
axisymmetric perturbations, the only nonvanishing component of the
metric has the form
\begin{eqnarray}
g_{t\varphi}=-a\left(\frac{2M}{r}-\frac{Q^2}{r^2}\right)\sin^2\theta
+2QE_0\left(r-\frac{Q^2}{r}+\frac{Q^4}{2Mr^2}\right),
\end{eqnarray}
where $E_0$ is the strength of the asymptotically uniform external
electric field and $a$ is a small angular momentum parameter. As
mentioned in \cite{BD}, this is not valid for arbitrarily large
$r$ since a uniform field in an asymptotically flat spacetime
would contain an infinite amount of energy. It is meaningful only
for $r$ satisfying the condition $\mid rE_0\mid\ll1$. Setting
$E_0=0$, the metric represents a slowly rotating Kerr-Newman black
hole. With $a=0$, it gives the perturbation describing the
Reissner-Nordstr\"{o}m black hole immersed in the weak,
asymptotically uniform electric field of strength $E_0$. Using the
formula (\ref{Lft1}), we may then read off the angular momentum
within a sphere of radius $r$ as
\begin{eqnarray}
{\cal L}(r)=-a\left(M-\frac{2Q^2}{3r}\right)
+QE_0\left(Q^2-\frac{r^2}{3}-\frac{2Q^4}{3Mr}\right).
\end{eqnarray}
The first contribution explicitly comes from the rotation of the
Kerr-Newman black hole and it would have vanished if we had
introduced the Reissner-Nordstr\"{o}m black hole (of $a=0$) from
the outset. The second contribution comes from the electromagnetic
fields of the system and vanishes only as the external electric
field disappears. That is, it appears because the hole is immersed
in the external electric field. When $a=0$, the angular momentum
within the horizon is given by
\begin{eqnarray}
{\cal L}(r_+)=
-\frac23QE_0(M^2-Q^2)\left(1+\sqrt{1-\frac{Q^2}{M^2}}\right),
\end{eqnarray}
which is not zero unless the black hole approaches its
extremality. This may originate from the electromagnetic fields
inside the horizon, but may be perceived as usual rotation when
viewed from the outside in accordance with the no-hair theorem. It
is determined only by the charge and mass of the hole and the
strength of the external field. It is thus tempting to conjecture
that the magnetically charged black hole immersed in an external
electric field should be constantly rotating with an angular
momentum determined by those quantities. Let's suppose that one
tries to stop the rotation of the hole by extracting its
rotational energy (for example, via the Penrose process or the
superradiant scattering) keeping the position of the charge fixed.
However, during the extraction process, energy must be constantly
supplied to the system through the external field when he tries to
keep its strength constant. The rotation is kept constant, because
the angular momentum is determined only by the charge and mass of
the hole and the strength of the external field. Thus, it may not
be realistic to consider a non-rotating magnetically charged black
hole immersed in an external electric field.

On the same line, it may be impractical to consider a non-rotating
magnetically charged black hole immersed in the electric field of
the charge. Therefore, we have to consider first the system of a
charge and a rotating magnetically charged black hole, and then
disentangle the effect due to the rotation of the hole itself
leaving only the electromagnetic angular momentum. As a way to
disentangle it from the angular momentum, we consider the angular
momentum obtained by performing the Komar integral at the horizon:
${\cal L}_H\equiv{\cal L}(r_H)$, and then drop the contribution
coming from this quantity to the total angular momentum measured
in the asymptotic region: $J\equiv{\cal L}(\infty)$. To do this,
we first consider the angular momentum of a slowly rotating
dilaton black hole and examine the relation between the two
quantities.

~\\
{\bf A. Angular Momentum of a Slowly Rotating Dilaton Black Hole}
~\\

We first recall the angular momentum of a slowly rotating dilaton
black hole mentioned in section \ref{II}. Inserting the metric
component $g_{t\varphi}$ of (\ref{srbh}) into the formula
(\ref{Lft1}) and (\ref{chi}), we get
\begin{eqnarray}\label{sram}
{\cal L}(r)=-\frac{a}{6}R^4(r)f'(r)
=\frac{a}{6}\left(3r_++\frac{3-\alpha^2}{1+\alpha^2}r_--
\frac{4}{1+\alpha^2}\frac{r_+r_-}{r}\right).
\end{eqnarray}
The angular momentum is directly read by
\begin{eqnarray}\label{amT}
J\equiv{\cal L}(\infty)=\frac{a}{6} \left(3r_++\frac{3-\alpha^2}
{1+\alpha^2}r_-\right),
\end{eqnarray}
which is coincident with the result obtained in \cite{HH}. Thanks
to the formula (\ref{sram}), we can write the angular momentum
contained between two spheres with radii $r_1$ and $r_2$ as the
difference: ${\cal L}(r_2)-{\cal L}(r_1)$. We can then divide the
angular momentum into two parts. The first part is the angular
momentum inside the horizon. It may be interpreted as the
intrinsic spin of the black hole and its magnitude is
\begin{eqnarray}\label{amH}
{\cal L}_H\equiv {\cal L}(r_+)=\frac{a}{6}\left(3r_+-r_-\right).
\end{eqnarray}
The second part is the angular momentum outside the horizon and is
written by
\begin{eqnarray}\label{amF}
{\cal L}_{out}\equiv {\cal L}(\infty)-{\cal L}(r_H)
=\frac{2r_-a}{3(1+\alpha^2)}=\frac{2}{3}\frac{Q^2a}{r_+}.
\end{eqnarray}
This part can be regarded as the angular momentum stored in the
electromagnetic field outside the black hole. Since there exists a
non zero component of electric field induced due to the rotation
of the black hole, the electromagnetic angular momentum is not
zero as it would be for a non-rotating charged or a rotating
neutral black hole.

Eliminating $a$ by using (\ref{amH}) from (\ref{amT}), the angular
momentum of a slowly rotating dilaton black hole can be rewritten
in terms of the horizon angular momentum as
\begin{eqnarray}
J=\frac{{\cal L}_H}{\left(3r_+-r_-\right)}
\left(3r_++\frac{3-\alpha^2}{1+\alpha^2}r_-\right).
\end{eqnarray}
We will use this relation to decouple the effect of the spin of
black hole from the electromagnetic field angular momentum below.

~\\
{\bf B. Angular Momentum stored in the Electrostatic and
Magnetostaic Fields}
~\\

Here we construct an expression for the electromagnetic angular
momentum using the properties, discussed in the previous section,
of the particular and the homogeneous solutions of the
second-order differential equation (\ref{sndeq}) in terms of
$\ell_1$, $\ell_2$, and $\ell_p$ and their differentials. From the
properties of the solutions, ${\cal L}(r)$ can be written as
\begin{eqnarray}
{\cal L}(r)=k_1\ell_1(r)+k_p \ell_p(r),~~~~r<b,\\
{\cal L}(r)=k_2\ell_2(r)+\tilde{k}_p \ell_p(r),~~~~r>b,\label{lk}
\end{eqnarray}
where $k_1$, $k_2$, $k_p$, and $\tilde{k}_p$ are coefficients to
be determined from matching conditions at $r=b$. The continuity of
${\cal L}$ and $d{\cal L}/dr$ at $r=b$ yields two conditions:
\begin{eqnarray}
k_1\ell_1(b)-k_2\ell_2(b)+\bar{k}_p\ell_p(b)=0,\\
k_1\ell_1^{\prime}(b)-k_2\ell_2^{\prime}(b)+\bar{k}_p\ell_p^{\prime}(b)=0,
\end{eqnarray}
where $\bar{k}_p\equiv k_p-\tilde{k}_p$~. The discontinuity of
${\cal DL}$ at $r=b$ yields that
\begin{eqnarray}
\bar{k}_p~{\cal D}\ell_p|_{r=b}=2eQ,
\end{eqnarray}
where we used the fact that $\ell_1$ and $\ell_2$ are solutions of the
homogeneous equation, {\it i.e.}, ${\cal D}\ell_1(r)={\cal D}\ell_2(r)=0$.
$k_1,~k_2,~\bar{k}_p$ are determined by above three conditions:
\begin{eqnarray}
k_1&=&-\frac{\ell_p(b)\ell_2^\prime(b)-\ell_p^\prime(b)\ell_2(b)}
{\ell_1(b)\ell_2^\prime(b)-\ell_1^\prime(b)\ell_2(b)}\bar{k}_p~,\label{k1}\\
k_2&=&-\frac{\ell_p(b)\ell_1^\prime(b)-\ell_p^\prime(b)\ell_1(b)}
{\ell_1(b)\ell_2^\prime(b)-\ell_1^\prime(b)\ell_2(b)}\bar{k}_p~,\label{k2}\\
\bar{k}_p&=&-\frac{6eQ}{a}\left(3r_+
+\frac{3-\alpha^2}{1+\alpha^2}r_-\right)^{-1}\label{kp},
\end{eqnarray}
where we used (\ref{ccc}) in the last line. From (\ref{lpr}) and
(\ref{lk}), it is straightforward to show that the total angular
momentum is expressed in term of $\tilde{k}_p$ as follows
\begin{eqnarray}
J=\lim_{r\rightarrow\infty}{\cal L}(r)
 =\tilde{k}_p\lim_{r\rightarrow\infty}\ell_p(r)
 =\frac{a}{6}\left(3r_++\frac{3-\alpha^2}{1+\alpha^2}r_-\right)
  \tilde{k}_p.
\end{eqnarray}
Therefore, we still need another additional condition to determine
$\tilde{k}_p$. We may think that the total angular momentum is
composed of two parts. The first part is the angular momentum due
to the rotation of the black hole, which includes the intrinsic
spin of the black hole and the contribution from magnetic and
electric fields induced due to its rotation. The second part is
the angular momentum stored in the electromagnetic field
unaffected by the rotation of the black hole. This part
corresponds to the angular momentum of an electric charge held at
rest around a non-rotating magnetically charged black hole that
would have if such configuration could been set up.

Since we are mainly interested in the latter, we separate out the
former from the total angular momentum. To do this, we suppose
that the angular momentum contained within the horizon is to be
${\cal L}_H={\cal L}(r_+)$, that is,
\begin{eqnarray}\label{lhk}
{\cal L}_H=k_1\ell_1(r_+)+k_p \ell_p(r_+).
\end{eqnarray}
The coefficient $\tilde{k}_p$ is determined from (\ref{k1}),
(\ref{kp}), (\ref{lhk}), and (\ref{lpr}) and the angular momentum
is given by the expression
\begin{eqnarray}
J=\frac{{\cal L}_H}{\left(3r_+-r_-\right)}
\left(3r_++\frac{3-\alpha^2}{1+\alpha^2}r_-\right)
+eQ\left(1-\frac{\ell_1(r_+)}{\ell_p(r_+)}
\frac{\ell_p(b)\ell_2^\prime(b)-\ell_p^\prime(b)\ell_2(b)}
{\ell_1(b)\ell_2^\prime(b)-\ell_1^\prime(b)\ell_2(b)}\right).
\end{eqnarray}
The first term of the right hand side represents the part due to
the rotation of the black hole as discussed in the previous
subsection. The second term is our main result of this subsection
and represents the angular momentum stored in the electromagnetic
field which are not affected by the spin of black hole. The
electromagnetic angular momentum $J_{em}$ can be rewritten as
follows
\begin{eqnarray}\label{mr}
J_{em}=eQ\left[1-\frac{\ell_1(r_+)}{\ell_p(r_+)}
\left(\frac{\ell_p(b)}{\ell_1(b)}-\frac{b^2\lambda^2(b)}{R^2(b)}
\left(\ell_p(b)\ell_1^\prime(b)-\ell_p^\prime(b)\ell_1(b) \right)
\frac{\ell_2(b)}{\ell_1(b)}\right)\right],
\end{eqnarray}
where we have used the relation (\ref{l12}) and eliminated
$\ell_2^\prime(b)$. Note that this quantity is independent of the
angular momentum parameter $a$, even though we introduced it into
the inhomogeneous solution $\ell_p$.

Unfortunately, exact forms of $\ell_1$ and $\ell_2$ are not known
for general coupling $\alpha$. We know explicit forms of them only
for $\alpha=0$ and $\sqrt{3}$. Therefore, we first examine
quantitatively the electromagnetic angular momentum for $\alpha=0$
and $\sqrt{3}$ with explicit forms of $\ell_1$ and $\ell_2$. We
then qualitatively estimate the behavior of $J_{em}$ using the
properties of $\ell_1$, $\ell_2$, and $\ell_p$ and confirm this
estimation by analyzing the behavior of ${\cal L}(r)$ for
arbitrary coupling $\alpha$. Finally, we will numerically find the
angular momentums for values $\alpha=1/\sqrt{3}$ and $1$. We
choose these values because the values $\alpha=$
$0,~1/\sqrt{3},~1, ~\sqrt{3}$ are well within regimes of $\alpha$
where the extremal black holes show qualitatively different
behaviors. Moreover, they are meaningful in the context of
supergravity and string theories. As mentioned in section
\ref{II}, when $\alpha=0$, the action (\ref{action}) is the
Einstein-Maxwell system with an uncoupled scalar, which we can
take to be constant. The theory with $\alpha=1$ arises naturally
in the reduction of the heterotic string on $T^6$ \cite{TO}. The
value $\sqrt{3}$ appears in Kaluza-Klein compactification from
five to four dimensions. Finally, when $\alpha=1/\sqrt{3}$, the
four-dimensional extreme dilaton black hole is interpreted as the
double-dimensional reduction of a nonsingular five-dimensional
black string which is a solution of the pure five-dimensional
supergravity \cite{GHT}.

~\\
{\bf C. For $\alpha=0$ }
~\\

When $\alpha=0$, two independent solutions of the homogeneous
equation are given by
\begin{eqnarray}
\ell_1(r)&=&r^2-3r_+r_-+\frac{4r_+^2r_-^2}{r_++r_-}~\frac1r,\\
\ell_2(r)&=&\ell_1(r)\int_r^\infty\frac{dr'}{\lambda^2(r')
\ell_1^2(r')}\nonumber \\
&=&-\frac{(r_++r_-)^2}{(r_+-r_-)^4}\left[r+\frac12(r_++r_-)
-\frac23\frac{r_+r_-(r_+^2+10r_+r_-+r_-^2)}{(r_++r_-)^2}
\frac{1}{r}\right]\nonumber \\
&&-\frac{(r_++r_-)^2}{(r_+-r_-)^5}\left[r^2-
3r_+r_-+\frac{4r_+^2r_-^2}{r_++r_-}\frac{1}{r}\right]
\ln\left(\frac{r-r_+}{r-r_-}\right),
\end{eqnarray}
and the inhomogeneous solution of Eq.(\ref{sndeq}) is given by
\begin{eqnarray}
\ell_p(r)=\frac{a}{6}\left(3(r_++r_-)-\frac{4r_+r_-}{r}\right),
\end{eqnarray}
with $C=-a(r_++r_-)$. We can easily check that all the properties
mentioned in the previous section hold with these explicit forms
of solutions. $\ell_1$ is finite at $r=r_+$ and
$\ell_1(r_+)=r_+(r_+-r_-)^2/(r_++r_-)$, while it is divergent
according to $r^2$ as $r\rightarrow \infty$. $\ell_2$ diverges in
the order of $\ln (r-r_+)$ at $r=r_+$, and it goes to zero
according to $r^{-1}$ as $r\rightarrow \infty$. On the other hand,
in the case of the extremal black hole, $\ell_1$ vanishes by order
$(r-r_+)^2$, while $\ell_2$ diverges by order $(r-r_+)^{-3}$ at
$r=r_+$.

When we insert these solutions into (\ref{mr}), the
electromagnetic angular momentum is calculated as follows
\begin{eqnarray}\label{mr0}
J_{em}=eQ\left[1-\frac{\ell_1(r_+)}{\ell_p(r_+)}
\left(\frac{\ell_p(b)}{\ell_1(b)}-a(r_++r_-)\left(b-2\frac{r_+r_-}{r_++r_-}\right)
\lambda^2(b)\frac{\ell_2(b)}{\ell_1(b)}\right)\right].
\end{eqnarray}
We can easily check that this result is coincident with the result
obtained for the Reissner-Nordstr\"{o}m black hole in \cite{SJR}.
The angular momentum $J_{em}$ changes from 0 to $eQ$ as $b$ goes
from $r_+$ to $\infty$. Furthermore, by differentiating
(\ref{mr0}) with respect to $b$ one can easily show that $J_{em}$
is a monotonic function of $b$. That is, $J_{em}$ changes
monotonically from 0 to $eQ$. In the case of extremal black holes,
$\ell_1(r_+)=0$ and the second term in the square bracket of
(\ref{mr0}) vanishes identically. The angular momentum is then
precisely $eQ$ regardless of the separation distance $b$ between
the charge and the black hole. Plots for $J_{em}$ of (\ref{mr0})
with respect to $b$ are shown in Figure 1. (a).

~\\
\begin{figure}[htbp]\label{plot1}
\centerline{
\begin{tabular}{cc}
\epsfig{file=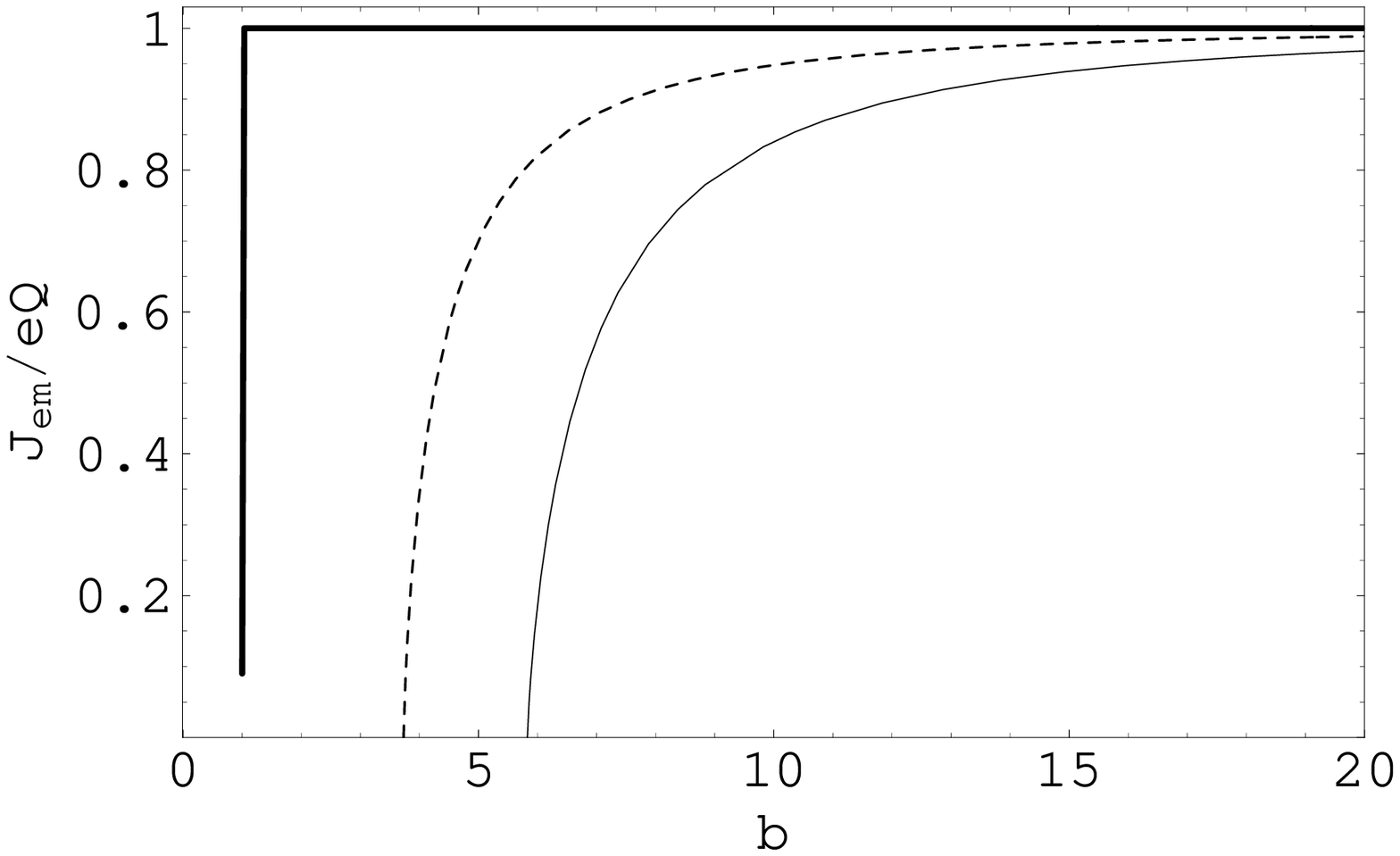, width=8cm, height=5.5cm} &
\epsfig{file=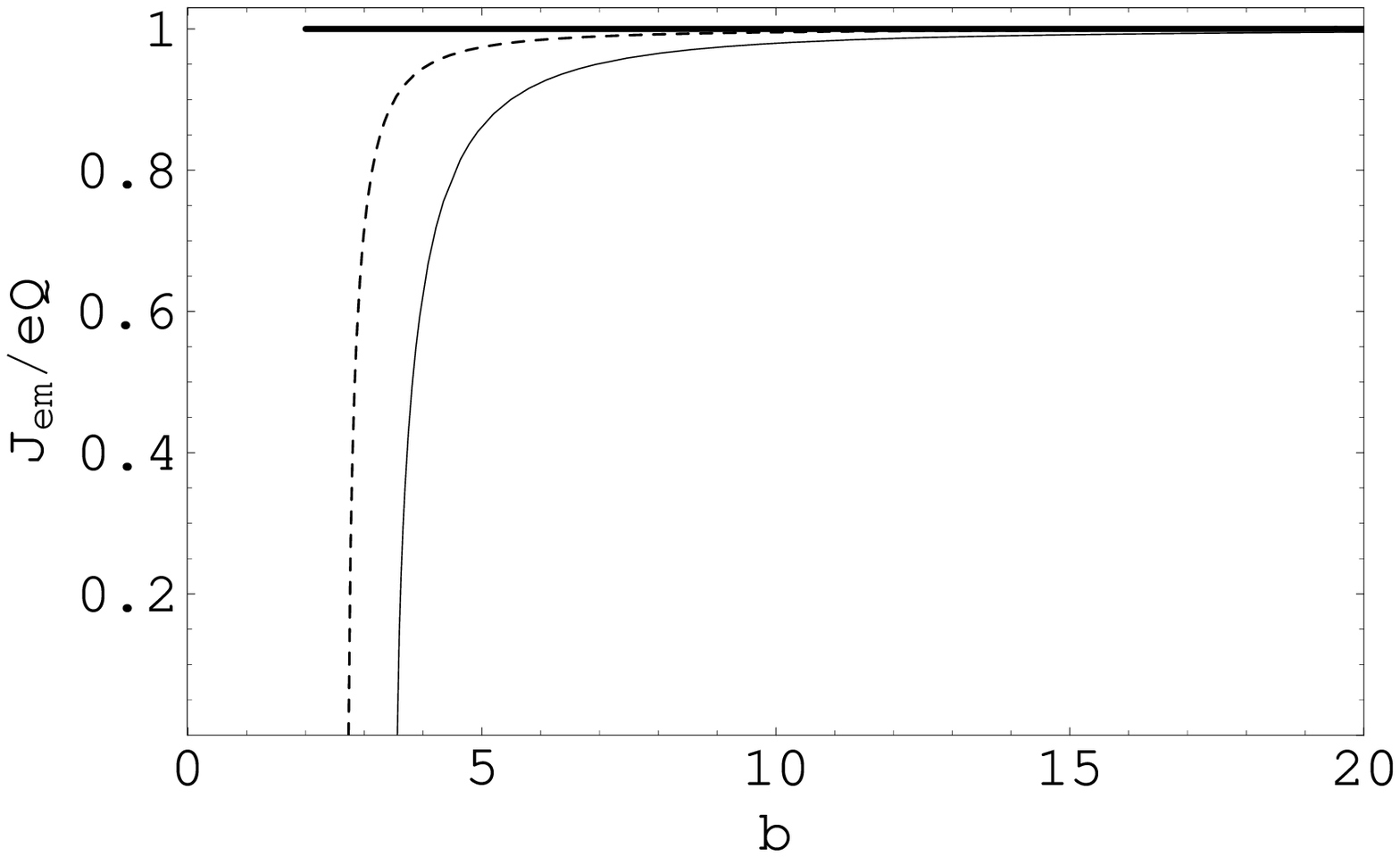, width=8cm, height=5.5cm} \\
$(a)~~~\alpha=0$ & $(b)~~~\alpha=\sqrt{3}$
\end{tabular}}
\caption{$J_{em}/eQ$ with respect to $b$ when $e=0.001$ and $Q=1$
for $\alpha=0$ and $\sqrt{3}$. The thick solid line is for an
extremal black hole with the critical mass
$M_c=Q/\sqrt{1+\alpha^2}$, the dashed line and the thin solid line
for a black hole with $M=2M_c$ and $3M_c$, respectively.}
\end{figure}

~\\
{\bf D. For $\alpha=\sqrt{3}$}
~\\

When $\alpha=\sqrt{3}$, two independent solutions of the
homogeneous equation are as follows:
\begin{eqnarray}
\ell_1(r)&=&\frac1r(r-r_-)^3, \\
\ell_2(r)&=&\ell_1(r)\int_r^\infty\frac{R^2(r')}{{r^\prime}^2
\lambda^2(r^\prime)\ell_1^2(r^\prime)} dr^\prime \nonumber \\
&=&-\frac{r_+}{(r_+-r_-)^3}\left[r+\frac12(r_+-5r_-)
-\frac{r_-}{6r_+}\left(r_+^2-5r_+r_--2r_-^2\right)
\frac{1}{r}\right]\nonumber \\
&&-\frac{r_+}{(r_+-r_-)^4}\frac{(r-r_-)^3}{r}
\ln\left(\frac{r-r_+}{r-r_-}\right),
\end{eqnarray}

The inhomogeneous solution follows from (\ref{lpr}) as
\begin{eqnarray}
\ell_p(r)=\frac{a}{6}\left( 3r_+-\frac{r_+r_-}{r}\right),
\end{eqnarray}
with a constant nonhomogeneous term $C=-a r_+$. Evidently, these
solutions also satisfy all the properties mentioned in the
previous section as for $\alpha=0$: $\ell_1(r_+)=(r_+-r_-)^3/r_+$
at $r=r_+$ and diverges according to $r^2$ as $r\rightarrow
\infty$. $\ell_2$ diverges in the order of $\ln (r-r_+)$ at
$r=r_+$, and it goes to zero according to $r^{-1}$ as
$r\rightarrow \infty$. On the other hand, in the case of the
extremal black holes, $\ell_1$ vanishes by order $(r-r_+)^3$,
while $\ell_2$ diverges by order $(r-r_+)^{-1}$ at $r=r_+$.

Inserting these solutions into (\ref{mr}), we obtain the following
expression for the electromagnetic angular momentum:
\begin{eqnarray}\label{mrs3}
J_{em}=eQ\left[1-\frac{\ell_1(r_+)}{\ell_p(r_+)}
\left(\frac{\ell_p(b)}{\ell_1(b)}
-ar_+(b-r_+)\frac{\ell_2(b)}{\ell_1(b)}\right)\right].
\end{eqnarray}
We can easily check that, as for $\alpha=0$, $J_{em}$ changes
monotonically from 0 to $eQ$ as $b$ goes from $r_+$ to $\infty$
for non-extremal black holes and is $eQ$ regardless of the
separation distance $b$ for extremal black holes. Plots for
$J_{em}$ of (\ref{mrs3}) with respect to $b$ are shown in Figure
1. (b).

~\\
{\bf E. For arbitrary $\alpha$}
~\\

Even though we don't have explicit forms of homogeneous solutions
$\ell_1$ and $\ell_2$, we can qualitatively estimate the
electromagnetic angular momentum from (\ref{mr}) using the
properties of the solutions discussed in the previous section.
First of all, as for the case $\alpha=0,~\sqrt{3}$, the angular
momentum depends upon the separation distance $b$ between the
black hole and the electric charge. Since $\ell_2$ converges by
order $r^{-1}$ and $\ell_1$ diverges by order $r^2$ as $r$
approaches $\infty$, the second term vanishes as the separation
distance goes to infinity. That is, (\ref{mr}) says that, as
$b\rightarrow\infty$, the electromagnetic angular momentum
approaches $eQ$. On the other hand, when the electric charge
approaches the horizon, the angular momentum becomes zero, {\it
i.e.}, $J_{em}\rightarrow 0$ as $b\rightarrow r_+$. This is easily
seen from observation that the second term in the large round
bracket vanishes as $b\rightarrow r_+$. It comes from the facts
that $\ell_1$ goes to a finite value, $\ell_2$ diverges only by
order $\ln (r-r_+)$, but $\lambda^2$ vanishes by order $(r-r_+)$
as $r\rightarrow r_+$. Thus the angular momentum $J_{em}$ changes
from $0$ to $eQ$ as $b$ goes from $r_+$ to $\infty$.

In the case of extremal black holes, {\it i.e.}, when $r_+=r_-$,
$\ell_1$ goes to zero by order larger than $(r-r_+)^2$ near the
horizon as mentioned above, and $\ell_1$ vanishes identically at
$r=r_+$. The second term in the square bracket then vanishes also.
Thus, the angular momentum is precisely $eQ$ regardless of the
separation $b$ between the charge and the dilaton black hole, as
for the monopole-charge system in flat space.

These properties are confirmed numerically. The electromagnetic
angular momenta $J_{em}/eQ$ are plotted versus the separation
distance $b$ in Figure 2. for extremal and non-extremal black
holes. The numerical results for $\alpha=1/\sqrt{3}$ and $1$ has
similar behavior as for $\alpha=0,~~\sqrt{3}$.

~\\
\begin{figure}[htbp]\label{abtalpha}
\centerline{
\begin{tabular}{cc}
\epsfig{file=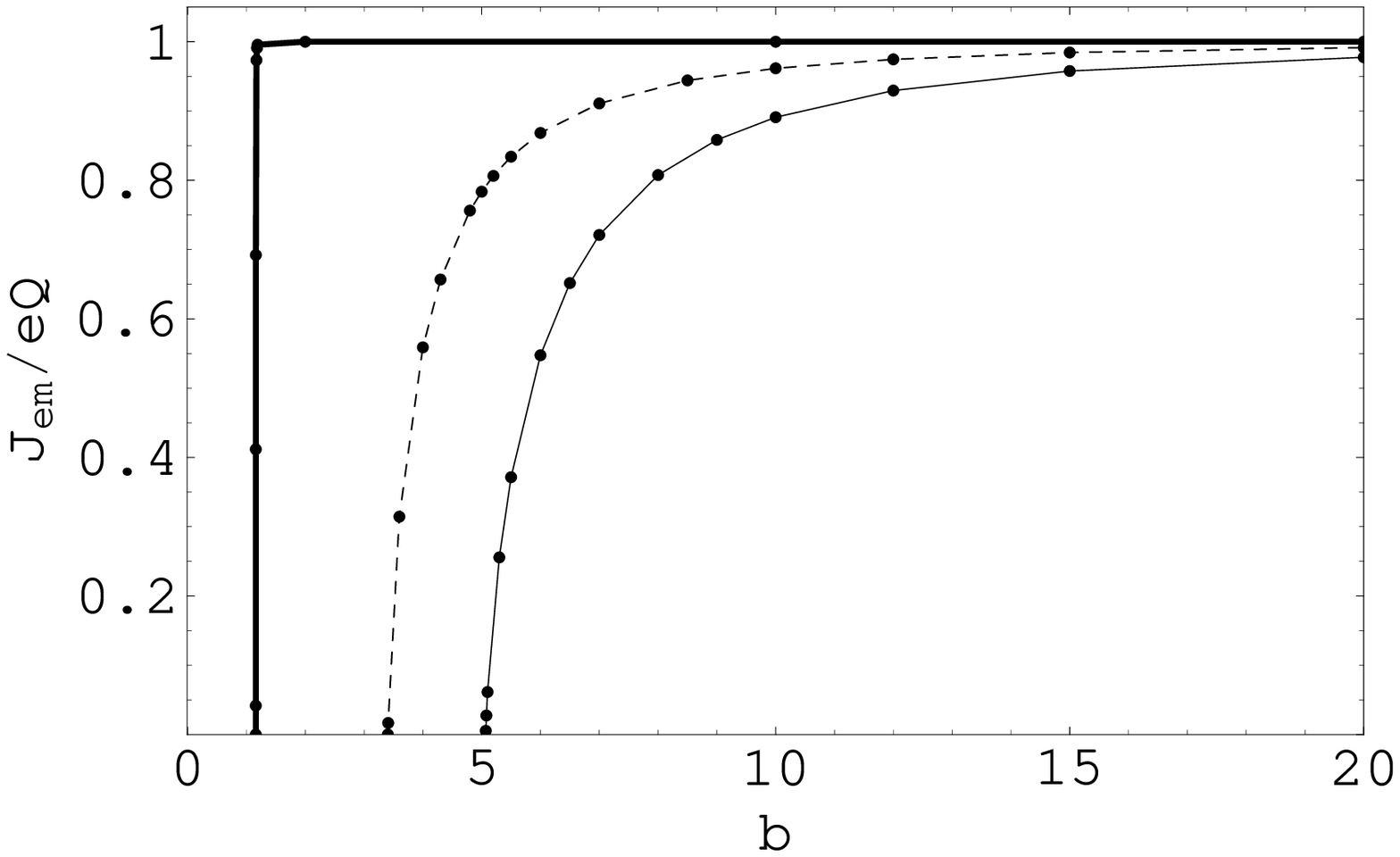, width=8cm, height=5.5cm} &
\epsfig{file=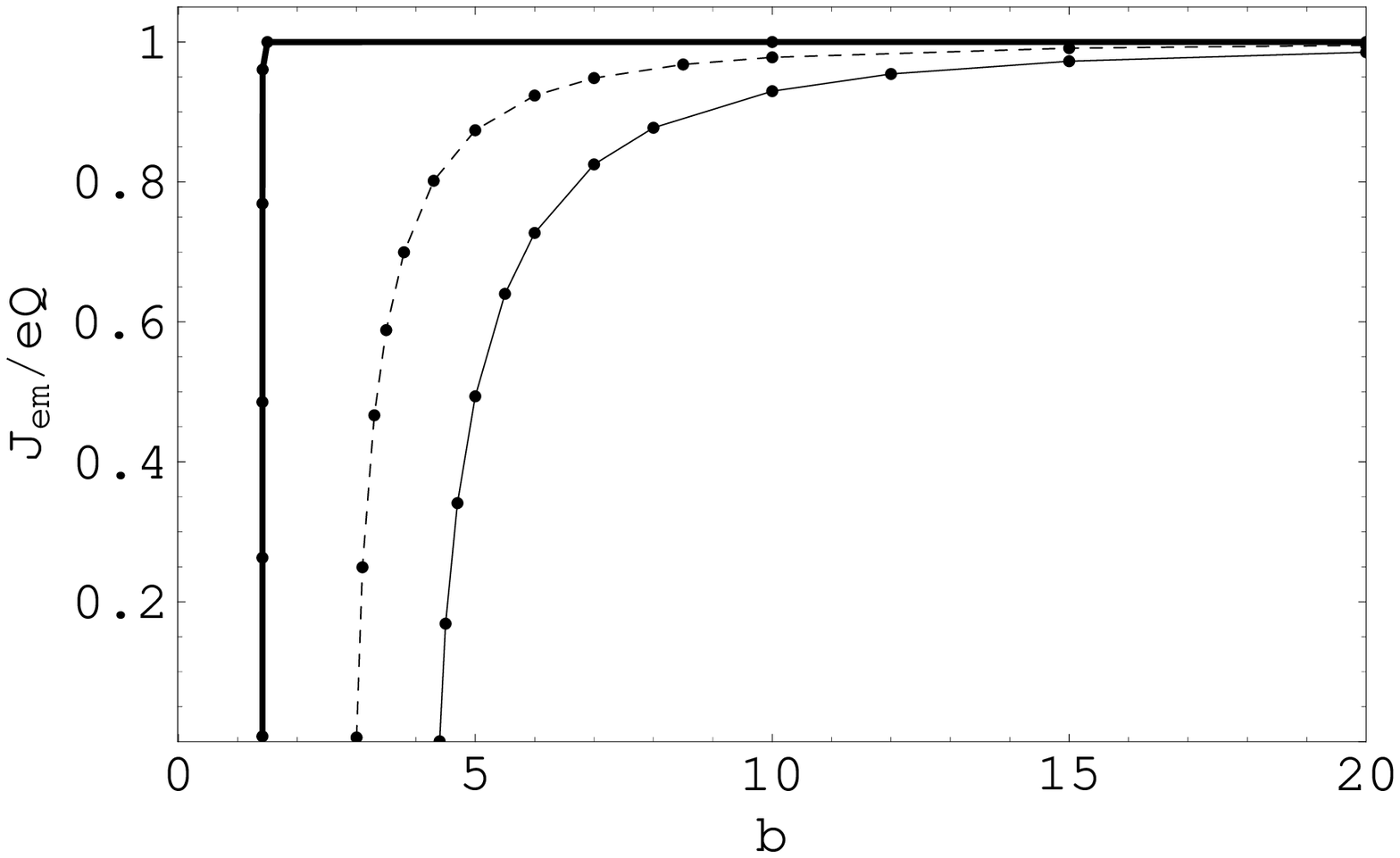, width=8cm, height=5.5cm} \\
$(a)~~~\alpha=1/\sqrt{3}$ & $(b)~~~\alpha=1$
\end{tabular}}
\caption{$J_{em}/eQ$ with respect to $b$ when $e=0.001$ and $Q=1$
for $\alpha=1/\sqrt{3}$ and $1$. The thick solid line is for an
extremal black hole with $M_c=Q/\sqrt{1+\alpha^2}$, the dashed
line and the thin solid line for a black hole with $M=2M_c$ and
$3M_c$, respectively.}
\end{figure}
~\\

So far we have examined the angular momentum of an electric charge
placed at rest outside a magnetically charged dilaton black hole.
We have seen that the electromagnetic angular momentum, the part
independent of the rotation of the black hole, shows several
common features regardless of the dilaton coupling strength, even
though the qualitative features of black holes are drastically
different depending on the dilaton coupling. It changes
monotonically from $eQ$ to $0$ as the charge goes down from
infinity to the horizon. However, in the extremal black hole case,
it is precisely $eQ$ as in the charge-monopole system in flat
space and its dependence upon the separation distance disappears.

Before ending this section, we would like to see what happens when
the charge approaches the horizon from infinity. Let's suppose the
black hole does not rotate at the outset and the charge is placed
at rest at an infinite distance from it. Then the angular momentum
of the system is all stored in the electromagnetic field outside
the horizon and its magnitude is simply $eQ$. As the charge moves
down radially toward the horizon, the black hole starts to rotate
and the spin angular momentum $J_s$ increases monotonically
because the total angular momentum $J$ is conserved, while the
electromagnetic angular momentum $J_{em}$ decreases monotonically.
When the charge reaches the horizon, the transfer is completed,
{\it i.e.}, $J_{em}$ is decreased in magnitude from $eQ$ to zero,
while $J_s$ is increased from zero to $eQ$. As a result, the
magnetically charged black hole turns into a dyon black hole
charged with $e$ and $Q$ and rotating with the total angular
momentum of $eQ$. On the other hand, for the extreme black hole
$J_{em}=eQ$ for all separation distance between the charge and
black hole. Thus as the charge moves down toward the black hole,
the angular momentum transfer does not occur and the black hole
does not spin up.

\section{Discussions}\label{VI}

We have analyzed the angular momentum of an electric charge placed
at rest outside a magnetically charged dilaton black hole. Even
though the qualitative features of black holes depend crucially on
the dilaton coupling, the angular momentum stored in the
electromagnetic fields shows several common features regardless of
the strength of dilaton coupling. It changes monotonically from
$eQ$ to $0$ as the charge goes down from infinity to the horizon.
Therefore, one can spin up a magnetically charged black hole by
throwing an electric charge radially into it. However, in the
extremal black hole case, its dependence upon the separation
disappears and the field angular momentum becomes independent of
the distance between the two objects. That is, it is precisely
$eQ$ as for the system of an electric charge and a magnetic
monopole in flat space, so spinning up the black hole is not
possible as long as the charge stays outside the horizon as
pointed out in \cite{SJR}. Explicitly, these results should
equally hold for a system of a magnetic monopole and an
electrically charged black hole by the electromagnetic duality.

Clearly these features raise several questions. Why does the field
angular momentum change as the separation varies? That is, why is
it $eQ$ as the charge rests at infinity? Why is it zero as the
charge reaches the horizon? Why does its magnitude remain frozen
with the value $eQ$ in the extremal limit of black holes
independently of the separation distance? Finally, can we expect
those things to hold for other systems including different
magnetically charged black holes in various theories with more
complicated field contents? It seems to be natural to ask this
question, because those phenomena are common to all dilaton black
holes which have drastically different features depending on the
dilaton coupling. Discussions on these questions are presented
below.

At first glance, one may think that changing of the
electromagnetic angular momentum upon the separation distance
occurs only in the presence of gravity, because the
electromagnetic angular momentum of the charge-monopole system is
given by the product of the two charges irrespective of the
separation distance in the absence of gravity. However, it is not
true. The exotic result in flat space comes from only a simple
consideration that both electric and magnetic charges are
point-like. The contribution of the electromagnetic field to the
angular momentum is directly calculated from
\begin{eqnarray}\label{jem}
\vec{J}_{em}=\int d^3x~\vec{r}\times (\vec{E}\times\vec{B}).
\end{eqnarray}
With the fields of point-like electric charge and monopole given
by $\vec{E}=e\vec{r'}/{r'}^3$ and $\vec{B}=Q\vec{r}/r^3$ with
$r'\equiv |\vec{r}-\vec{b}|$ respectively, the angular momentum
density, integrand of (\ref{jem}), varies according to the
separation distance between the two objects, but the integral of
the density over the whole region is independent of the separation
distance. There is no way to transfer the electromagnetic angular
momentum to anything else. However, if we introduce a magnetically
charged classical object (with a nontrivial magnetic charge
distribution) instead of the point-like monopole, then we may
observe the separation dependence of the electromagnetic angular
momentum even in flat space.

To see this, we address a simple example in flat space: a system
of an electric charge and a magnetically charged non-conducting
spherical shell of radius $R$. The magnetic charges are evenly
distributed on the shell and the electric charge is located at a
distance $b$ apart from the center of the shell. The contribution
to the integral comes from only fields outside the shell because
the magnetic field vanishes inside the shell. When both objects
are at rest, the electromagnetic angular momentum is then directly
calculated to give
\begin{eqnarray}
J_{em}&=& eQ\left(1-\frac{R^2}{3b^2}\right)~~~{\rm for}~~b\geq R,
      \label{jemso}\\
      &=& eQ\frac{2b}{3R}~~~~~~~~~~~~~~{\rm for}~~b<R
      \label{jemsi}.
\end{eqnarray}
Certainly, the magnitude of the electromganetic angular momentum
varies with the separation distance: it changes from $0$ to $eQ$
as the electric charge moves from the center to infinity. This can
be explained as follows. When the charge is placed at infinity
from the shell, {\it i.e.}, when $b \gg R$, the electric field
flux across it vanishes. This is the same as letting $R\rightarrow
0$ while keeping $b$ finite. Then the angular momentum is $eQ$ and
is all stored in the electromagnetic fields outside the horizon.
When the electric charge is closer to the shell, the electric
field flux crossing the shell is larger. Then, the integration
(\ref{jem}) is smaller because the integral is restricted outside
the shell, unlike the charge and monopole system. Finally, when
the electric charge reaches the center of the shell, the density
$\vec{r}\times(\vec{E}\times\vec{B})$ becomes zero over the whole
region because both fields are purely radial from the center of
the shell. Of course, in a dynamical process the total angular
momentum should be conserved. As the electric charge moves toward
the shell, the latter is forced to rotate about its axis due to
the Lorentz force exerted on magnetic charges by the electric
field crossing it, and hence becomes spinning. Once the shell
starts to rotate, an electric dipole field is induced around it
due to magnetic charges rotating about the symmetric axis and a
part of the angular momentum is stored in this dipole field and
the existing magnetic field. That is, as the electric charge gets
closer, the electromagnetic angular momentum is transferred to the
spin of the shell and stored in the magnetic field and the induced
electric dipole field. The transfer becomes complete as the
electric charge reaches the center of the sphere. On the other
hand, the electromagnetic angular momentum becomes independent of
the separation distance as the shell shrinks to a point, {\it
i.e}, as $R\rightarrow 0$. This confirms that the change of the
electromagnetic angular momentum upon the separation distance
comes from the nontrivial magnetic charge distribution.

Another interesting example is a system of an electric charge and
a neutral conducting sphere in which a magnetic monopole is
centered. In the presence of the charge, two image charges can be
introduced within the conducting sphere, the first
$q^{\prime}=e(R/b)$ at the center and the second
$q^{\prime\prime}=-e(R/b)$ at a distance $b^{\prime\prime}=R^2/b$
from the center. The first image charge does not contribute to the
angular momentum because its electric field is purely radial like
the magnetic field of the monopole. The second contributes
negatively because its charge is opposite to that of the real
charge. The image charge is larger as the charge is closer to the
conducting sphere. When the electric charge contacts the sphere
finally, the second image charge becomes $-e$ and offsets the real
charge leaving only the first image charge of $e$ at the center.
Then the electromagnetic angular momentum vanishes because both
electric and magnetic fields are purely radial. On the contrary,
when the charge goes to infinity from the sphere, the image
charges are diminished and the electric field flux crossing the
sphere becomes negligible, and so the electromagnetic angular
momentum is all stored in fields outside the sphere and has the
magnitude of $eQ$. Taking into account the contribution of image
charges and using the results of the previous example, we obtain
the electromagnetic angular momentum for this system
\begin{eqnarray}\label{jemsii}
J_{em}=eQ\left(1-\frac{R^2}{b^2}\right)~~~{\rm for}~~b\geq R,
\end{eqnarray}
which, as expected, runs monotonically from $0$ to $eQ$ as the
electric charge goes from the sphere to infinity. This is very
similar to the result obtained for the system of an electric
charge and a magnetically charged black hole. Again, as
$R\rightarrow 0$, the electromagnetic angular momentum becomes
independent of the separation distance, showing that its change
upon the separation comes from the nontrivial electromagnetic
structure surrounding the monopole.

Let's return to the charge and black hole system. The result
obtained in the previous sections for the system can be explained
analogously to the above examples. In fact, the horizon of a black
hole plays a role very similar to that of the conducting sphere in
the above example, even if a black hole has no material surface
that differs from the surrounding space in contrast to the
conducting sphere. As is well known in the black hole
electrodynamics, for a distant observer the horizon can be thought
of as a fictitious surface having an electric charge density that
compensates for the flux of electric field across the horizon, and
an electric current that closes tangential components of the
magnetic fields at the horizon \cite{NF}. Thus, the separation
dependence of the electromagnetic angular momentum of an electric
charge and a magnetically charged black hole could be explained in
an analogous way to the above example introducing a charge
distribution on or image charge within the horizon, though the
electromagnetic fields are distorted and more complicated due to
the gravity around the black hole.

This can be interpreted from another perspective. In the previous
sections we have obtained the monotonic change of the angular
momentum upon the separation, only solving the equations of motion
(\ref{eom1})-(\ref{eom3}) without introducing {\it ad hoc}
electromagnetic properties of the horizon. Even though a black
hole simply has a point source at the origin and has no material
structure, it has a nontrivial spacetime structure and is shielded
by the event horizon. The horizon does such a role as the
conducting shell of the above example does. Of course, there are
no real material surface and no real charge distribution at the
horizon. However, the event horizon makes the electromagnetic
angular momentum across it indistinguishable from the intrinsic
spin of the black hole in accordance with the no hair theorem.

What happens is as follows. When the charge is placed at rest at
infinity from a non-rotating charged black hole, no flux of the
electric field passes through the black hole and so no flux of the
electromagnetic angular momentum crosses it. The integral of
(\ref{Lft}) on the horizon, ${\cal L}(r_+)$, then vanishes. The
angular momentum is all stored in the electromagnetic fields
outside the horizon and its value is then $eQ$ as for the charge
and monopole system in flat space.

On the other hand, when the charge approaches the black hole, the
electric field lines passing through the black hole increases and
also the flux of the electromagnetic field angular momentum across
the horizon increases. Once the electromagnetic angular momentum
crosses over the horizon, it becomes indistinguishable from the
spin of the black hole according to the no hair theorem. That
means the influx of the electromagnetic angular momentum across
the horizon appears as the intrinsic spin of the black hole. As
the electric charge gets closer to the black hole, the flux of the
electromagnetic angular momentum across the horizon increases and
the spin of the black hole becomes larger. If the black hole
rotates, the existing magnetic and electric fields are
supplemented with induced electric and magnetic fields
respectively, due to the dragging of the inertial frame into the
rotational motion around the black hole. These induced fields also
store an amount of electromagnetic angular momentum, which depends
on the rotation of the black hole and vanishes only when the
angular momentum parameter $a$ of the black hole becomes zero.
Therefore, as the charge approaches the black hole, the
electromagnetic angular momentum in the existing fields is
continuously transferred to the spin angular momentum and the
electromagnetic angular momentum in the induced fields.

When the electric charge reaches the horizon, all the
electrostatic multipoles except the monopole fade away in
accordance with the no hair conjecture. The electrostatic monopole
field due to this charge then becomes purely radial like the
magnetic monopole field of the black hole. So the angular momentum
stored in these monopole fields is zero and the transfer gets
complete. A part of it is transferred to the spin of the black
hole and the rest is stored in the induced fields. Once the charge
reaches the horizon, then the exterior fields are exactly those of
the rotating dyonic dilaton black hole with both the electric and
magnetic charges. For a slowly rotating dyonic dilaton black hole,
the metric can be displayed by (\ref{srbh}) replacing $Q^2$ with
$Q^2+e^2$ \cite{MK}. Since the total angular momentum is $eQ$, the
angular momentum parameter $a$ then can be written, from
(\ref{srbham}), as
\begin{eqnarray}
a=6eQ\left(
3\bar{r}_++\frac{3-\alpha^2}{1+\alpha^2}\bar{r}_-\right)^{-1},
\end{eqnarray}
where $\bar{r}_\pm$ are positions of outer and inner horizons of
the dyonic dilaton black hole and are obtained from (\ref{mpm})
and (\ref{qpm}) by replacing $Q^2$ with $Q^2+e^2$. Then, from
(\ref{amH}) and (\ref{amF}), the angular momentum within the
horizon is given by
\begin{eqnarray}
{\cal L}_H=eQ\frac{(1+\alpha^2)(3\bar{r}_+-\bar{r}_-)}
{3(1+\alpha^2)\bar{r}_+ +(3-\alpha^2)\bar{r}_-},
\end{eqnarray}
and the angular momentum in the electromagnetic field induced by
the rotation of the black hole is
\begin{eqnarray}
{\cal L}_{out}=
eQ\frac{4\bar{r}_-}{3(1+\alpha^2)\bar{r}_++(3-\alpha^2)\bar{r}_-}.
\end{eqnarray}
From above arguments, we can conclude that the separation
dependence of the electromagnetic angular momentum of the charge
and black hole system originates from the existence of the
horizon, which serves as a nontrivial spacetime structure and
obliterates all details of the angular momentum swallowed up in
accordance with the no hair theorem.

Finally, let's see what happens as the black hole is extremal from
the outset. The results of previous sections say that the angular
momentum stored in the electromagnetic field outside the horizon
is precisely $eQ$ and is independent of the separation distance
between the charge and the black hole. It seems to be easy to give
an explanation on this phenomenon for the dilaton black hole. As
mentioned in section \ref{II}, when $\alpha> 0$, the proper size
of the extremal black hole is zero in that the area of the horizon
$4\pi R^2(r_+)$ vanishes and the degenerate horizon at $r=r_+$ is
singular. The extremal black hole looks like a point-like object.
The situation is then the same as with the charge-monopole system
in flat space. The electric field flux passing the black hole is
zero. The angular momentum is all stored in the electromagnetic
field outside the horizon and has the constant value $eQ$.

This kind of explanation does not seem to hold for the system of
the electric charge and the extremal Reissner-Nordstr\"{o}m black
hole (as $\alpha=0$), because its horizon area is $4\pi r_+^2$ and
its proper size is not zero. We may then naively expect that its
cross-sectional area for the capture of the electric flux should
be $\pi r_+^2$ and the electric flux across the hole should be
non-vanishing. According to our naive expectation, the
contribution to the total angular momentum of the system from
behind the horizon should be non-zero in the extremal limit in
this configuration. However, this contradicts the above result
that the electromagnetic angular momentum outside the
Reissner-Nordstr\"{o}m black hole is precisely $eQ$, independently
of the separation distance between the charge and the black hole.

The puzzle is solved when we invoke that the component of the
electromagnetic fields normal to the horizon tends to be zero and
the flux lines are totally expelled, as the black hole approaches
extremality. This is analogous to the ``Meissner effect" for a
magnetic field around a superconductor. This phenomenon was first
pointed out by J. Bi$\check{\rm c}\acute{\rm a}$k and
Dbo$\check{\rm r}\acute{\rm a}$k \cite{BD} in the context of the
Einstein-Maxwell theory. They considered the electrically charged
Reissner-Nordstr\"{o}m black hole immersed in the weak,
asymptotically uniform magnetic field and found that the magnetic
flux lines are expelled from the black hole in the extremal limit.
Later, it was shown by Chamblin, Emparan and Gibbons \cite{CEG}
that this effect broadly exists for various extremal black hole
solutions including the dilaton black hole and other extremal
solitonic objects (such as $p$-branes) in string theory and
Kaluza-Klein theory.

Due to the electromagnetic duality, we can expect that the
electric field is expelled from the magnetically charged black
hole in the extremal limit. Then the effective cross-sectional
area for the capture of the electric flux shrinks to zero and the
electric flux across the hole vanishes. The contribution to the
total angular momentum of the system from behind the horizon is
then zero, because the black hole is non-rotating from the outset
and the electric field inside the horizon does not exist. Thus,
even though the proper size of the horizon is not zero in the
extremal limit, the angular momentum is all stored in the
electromagnetic field outside the horizon as in the case of the
charge-monopole system in flat space and is precisely $eQ$.

On the other hand, we could give an analogous explanation to the
system of the charge and the extremal dilaton black hole (as
$\alpha>0$). As is well known, in four dimensions the gauge field
equation is conformally invariant. This means that, when we deal
with the dilaton black hole, the electromagnetic field does not
distinguish whether we are working in the Einstein frame or any
other conformal frame related to it by an overall rescaling of the
metric by a factor of the dilaton as mentioned in \cite{CEG}. In
this sense, we could just as well work in a conformally related
metric where the extremal horizon is non-singular. From the
dilaton black hole solution (\ref{emf})-(\ref{dbh}), we easily see
that such a conformal frame is $e^{2\alpha\phi}ds^2$. In this
frame the extremal black hole area is equal to $4\pi r_+^2$.
However, the electric flux across the horizon vanishes as shown in
\cite{CEG}. Once again, for the system of the charge and the
extremal dilaton black hole the contribution to the angular
momentum from inside the horizon is zero, because the black hole
is non-rotating from the outset and the electric field is absent
inside the horizon. As a result, the field flux expulsion from the
extremal black hole horizon leads to the phenomenon that the
electromagnetic angular momentum of the system comprising an
electric charge and an extremal magnetically charged black hole is
independent of the separation distance between the two objects and
precisely $eQ$ as for the charge-monopole system in flat space.

In conclusion, above arguments appear to settle all the questions
in the second paragraph of this section. The reason why the
electromagnetic angular momentum changes from $eQ$ to $0$ as the
charge is brought down from infinity to the horizon is due to the
no-hair nature of black holes in that the horizon makes the
electromagnetic field angular momentum across it indistinguishable
from the intrinsic spin of the black hole and it makes the
electric field purely radial (leaving only the monopole field)
when the charge reaches the horizon. Next, the electromagnetic
angular momentum having the precise value $eQ$ in the extremal
limit of black holes independently of the separation distance is
essentially related to the phenomenon that the electric flux lines
are expelled from extremal black holes. Since both the no-hair
nature of black holes and the field expulsion from extremal black
holes are generic for black holes, we expect aforementioned
properties of the electromagnetic angular momentum to be generic
for systems of an electric charge and a magnetically charged black
hole (or a magnetic monopole and an electrically charged black
hole, by the electromagnetic duality) in various theories with
more complicated field contents, namely, in string theory and
Kaluza-Klein theory.

~\\
{\Large {\bf Acknowledgments}} \\
~\\
S.H.M would like to thank Soo-Jong Rey for introducing him this
topic many years ago. J.H.K. is supported by the Korea Research
Foundation Grant funded by the Korean Government (MOEHRD)(Grant
R14-2003-012-01002-0).



\end{document}